\documentclass[multphys,vecphys]{svmult}

\usepackage{makeidx}         
\usepackage{graphicx}        
\usepackage{multicol}        
\usepackage[bottom]{footmisc}
\usepackage[latin1]{inputenc}
\usepackage{natbib}
\usepackage{journals}

\makeindex             

\begin{document}

\title*{Integral Field Spectrographs: a user's view}
\author{Eric Emsellem\inst{1}}
\institute{Université de Lyon 1, Centre de Recherche Astrophysique de Lyon, Observatoire
de Lyon, 9 avenue Charles André, 69230 Saint-Genis Laval, France ; CNRS,
UMR 5574 ; ENS de Lyon, Lyon, France.
\texttt{emsellem@obs.univ-lyon1.fr}
}
%
%
\maketitle

\section{Introduction}
\label{sec:intro}

We easily tend to think of Integral-Field Spectrographs (IFS) along two opposing
trends: as either the beautiful combination between photometry and spectroscopy,
or as our worst nightmare including the dark side of both worlds. I favour a
view where each IFS is considered individually, as one instrument with
specific performances which can be used optimally for a certain range of
scientific programs. It is indeed true that data-wise, IFS do sometime merge
the characteristics of classic (e.g., long-slit) spectrographs with annoying
issues associated with Imagers. This is in fact the price to pay to access a
drastically different perspective of our favourite targets. The challenge is
then to provide the necessary tools to properly handle the corresponding
data. However, this should certainly not be thought as something specific to
IFS: such a challenge should be accepted for any instrument, and most
importantly solved prior to its delivery at the telescope.

\section{Specifics}
\label{sec:spec}

Aperture and long-slit spectroscopy obviously provide a limited access
to the spatial information. The spectral and spatial characteristics
being mixed in the datasets (via e.g. the effect of seeing), this
can sometimes severely restrict our interpretations. By acceding to
the third dimension, we open the door to a wide variety of powerful
treatments such as : a posteriori detailed evaluation and correction 
of seeing effects, atmospheric diffraction, or testing the presence of 
artifacts (e.g. cosmic ray impacts). 

With IFS such as VIMOS (http://www.eso.org/instruments/vimos), we reached a rather
impressive multiplex level, with the possibility of acquiring more than 6000 spectra covering
more than 500 wavelength pixels. We are therefore nowadays reaching the limit of what we can
do in a non fully automated way: defects or singularities can still be detected via a manual
search or visualisation of individual (or group of) spectra, we can repeat complex
analysis process many times without worrying (too much) about the CPU cost, and
processing algorithms are allowed to crash once in a while without endangering the project.
When instruments such as MUSE will start gathering photons at the VLT (http://www.eso.org/instruments/muse), 
we will collect the equivalent of more than 100 VIMOS exposures in one single shot. 
For most scientific programs, there is therefore no way we will be able to
systematically look at individual spectra. We will then need robust algorithms to
analyse and process our data. More importantly, this implies that we should acknowledge
the presence of errors in our results, which should be readily evaluated. 

\section{Pushing the limits}
\label{sec:limits}

As mentioned above, there is a strong need for analysis tools in the context of IFS data:
basic software including slicing and visualisation techniques,
data mining are still not felt as standard, easy tasks, 
even though there exists a number of successful (but not universal) 
pieces of software. The data format is one obstacle, only partly answered by noticeable
efforts such as the one from the Euro3D consortium (see e.g., http://www.aip.de/Euro3D). 
We must now realise this situation may soon become untractable in a rather near future
if we do not prepare for it.
We then need to invent new or refine existing 
analysis tools to perform mosaicing, binning \citep{CapCop03}, optimal summation, 
normalisation, smoothing (spatially and spectrally), deconvolution, in 3D.
We could for example think of a drizzling \citep{Hook+00}
technique generalised for 3D data.

\begin{figure}
\centering
\includegraphics[width=11.5cm]{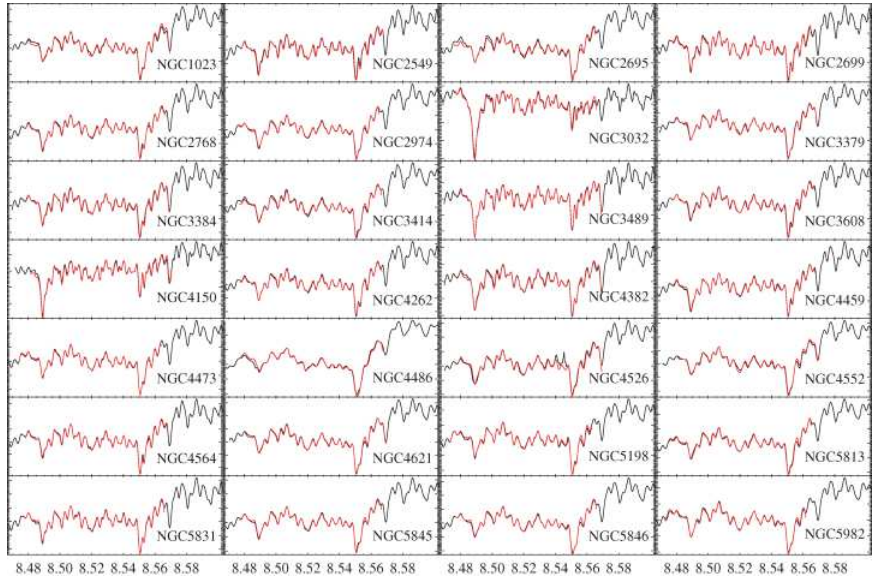}
\caption{Comparison of OASIS (in red) and SAURON (in black) optical spectra of the central
regions of galaxies, showing an impressive agreement when specifics of each instrument are 
properly taken into account. Extracted from \cite{McDermid+06}.}
\label{fig:OASIS}
\end{figure}
With the full 3D information in hand, we can proceed 
with efficient mosaicing procedures, both spatially or spectrally 
(see Fig.~\ref{fig:OASIS}), the optimisation of deep fields by taking into account
the specific characteristics (resolution varying over the field, signal-to-noise,
etc) of each individual exposure, the use of super-resolution \citep[see e.g.][]{Garcia+99},
or at last obtaining reasonable spectrophotometry. However, pushing the limits
is also demanding. If the spatial two-dimensional ingredient of such data
implies that models may then become much more constrained \citep{Cappellari+05}, it
also means that the uncertainties in the output scientific results
will even more critically depend on the errors in the dataset.

This is often, and wrongly in my opinion, seen by many as opening Pandora's box.
On the contrary, it should represent an additional motivation for a robust
estimate and {\em control} of such errors.
One dangerous asset of IFS data is that they can be used to build two-dimensional maps
of various quantities (e.g., emission line flux, kinematics, stellar
population indicators), and in general these maps look GOOD. These nice reconstructed
images are impressive but may hide severe defects in the datasets impairing
the subsequent scientific interpretation. It is therefore of the utmost
importance that we systematically evaluate errors associated with the
signal we wish to analyse. 

Noise must be propagated so that the user can keep track of it. An accurate assessment
of the noise is required for any optimal stacking, binning, etc.
Published measurement should always include errors bars, which in turn
depend on our ability to estimate the noise pattern in the data. 
Formats such as the Euro3D files do include a description of the noise
(per pixel). However, it can actually only keep track of the variance (the trace
of the noise matrix). Ignoring the noise covariance may sometimes
be unavoidable, but this comes at a cost which should be, again, evaluated
and taken into account.

\section{Implications and the fear of resampling}
\label{sec:implic}

\begin{figure}
\centering
\includegraphics[width=11.5cm]{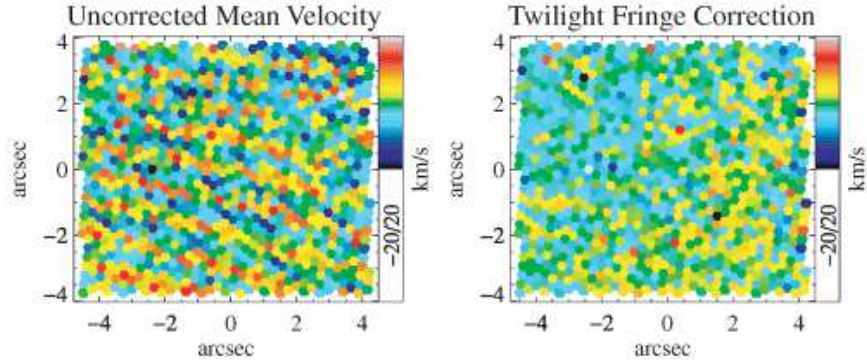}
\caption{Effect of fringe pattern as seen in the stellar velocity field of
a twilight exposure (left): the "zebra" pattern disappears after proper correction
of that effect. Extracted from \cite{McDermid+06}.}
\label{fig:zebras}
\end{figure}

We therefore need a better characterisation of our data. This calls for
a good, validated calibration system, with a detailed assessment of
the stability of the instrument and telescope. It also means extra caution
when e.g., taking into account the contamination by stray light, and more
importantly the thorough characterisation of the detectors. Subtle effects
such as very low uncorrected fringes may significantly impair the science
we wish to conduct (see Fig.~\ref{fig:zebras}).

The Data Reduction and Analysis pipelines represent another source of concern: 
they for example usually contain a number of steps
which require a resampling of the dataset (e.g. wavelength calibration). 
Resampling is often seen as evil, because it may spread artifacts over
several spaxels, and makes it difficult to follow the noise pattern.
Indeed, resampling means that pixels, spaxels, spectra, may not be
independent from each others anymore (as illustrated e.g., in the noise covariance).
The spreading of artifacts also dilutes the unwanted signal, making it less
likely to be detected. 

Spatial resampling can usually and should be avoided. We should therefore
develop an approach when we minimise the number of steps including some resampling.
This is naturally done for example when we apply the correction for the trace
distortions and the wavelength calibration at the same time.
One way out is to always keep working with the detector pixels,
and design the reduction and analysis steps via a global model approach
(deriving transformation matrices for each step and combining them
 before applying them). This is certainly a nightmare for
the software development people (and one in 3D!), but maybe not
an impossible task specially for densely-packed (fiber) systems
and image slicers. The other implication of such an approach is, again,
to consider the data analysis tools as closely {\em associated}
with the data reduction software, something we are for example 
seriously pushing in the context of the MUSE project.

We must finally redefine carefully what we do wish to call a science data product. 
Is flux calibration a requirement for all such datasets, and should flux
calibration be a procedure applied systematically? This obviously comes
with an additional operational cost, and should again be discussed
long before the first mounting of the instrument behind the telescope,
and not a posteriori when we realise we lack both the
understanding of our instrument, and the data for that specific 
calibration.

\section{Conclusions and perspectives}
\label{sec:conc}

From this very brief overview, we can already draw a number of conclusive
statements which may be important in the context of calibration procedures for
IFS data. Critical needs include:

\begin{itemize}
\item The development of the (automated ?) tools to analyse the huge data sets 
of the coming generation of instruments.
\item Keeping track of noise and systematic errors (a tricky task, but good for our scientific health).
\item A proper characterisation of the instrument (including the reduction software which must be
      seen as {\em part} of the instrument).
\item An adapted calibration plan, and the motivation for researchers to request telescope time
for {\em Calibration Proposals}.
\item The development of a good (parametric ?) instrument model.
\end{itemize}
In fact, most of these statements are absolutely not specific to IFS and should be applied
more generally to most instrument. Two final notes regarding software. Firstly, it is important
to realise that the software should be tested on (and developed with) realistic data. This
obviously requires the development of a good Instrumental Numerical Model, an approach
which already exists for space instruments, and seems to become the rule for
ground-based ones. In that context, the Data Reduction Software (DRS) and the Data Analysis Tools (DAT)
should be considered/designed in a consistent way so that our understanding of
the instrument and the DRS is also reflected in the optimisation of the DAT.
Secondly, there is clearly a need for more coordination in the development and follow-up upgrading
of data reduction softwares. As a user, I do not want to face the fact that several complete 
(and significantly different) versions of a pipeline exist in different places, 
specially when the best routines are privately owned. This requires that a strategic plan
exists for the maintenance of the software, and that the time-scales for the (official)
implementation of new recipes are short enough to avoid recurrent frustration on the user's side. 

I would like to warmly thank Pierre Ferruit for fruitful discussions, and the SOC for inviting me 
to this constructive Workshop.

\bibliographystyle{astron}
\bibliography{emsellem}

\begin{thebibliography}{}

\bibitem[\protect\astroncite{{Cappellari} and {Copin}}{2003}]{CapCop03}
{Cappellari}, M. and {Copin}, Y.: 2003,
\newblock {\em \mnras} {\bf 342}, 345

\bibitem[\protect\astroncite{{Cappellari} and {McDermid}}{2005}]{Cappellari+05}
{Cappellari}, M. and {McDermid}, R.~M.: 2005,
\newblock {\em Classical and Quantum Gravity} {\bf 22}, 347

\bibitem[\protect\astroncite{{Garcia} et~al.}{1999}]{Garcia+99}
{Garcia}, P.~J.~V., {Thi{\'e}baut}, E., and {Bacon}, R.: 1999,
\newblock {\em \aap} {\bf 346}, 892

\bibitem[\protect\astroncite{{Hook} and {Fruchter}}{2000}]{Hook+00}
{Hook}, R.~N. and {Fruchter}, A.~S.: 2000,
\newblock in N. {Manset}, C. {Veillet}, and D. {Crabtree} (eds.), {\em
  Astronomical Data Analysis Software and Systems IX}, Vol. 216 of {\em
  Astronomical Society of the Pacific Conference Series}, pp 521--+

\bibitem[\protect\astroncite{{McDermid} et~al.}{2006}]{McDermid+06}
{McDermid}, R.~M., {Emsellem}, E., {Shapiro}, K.~L., {Bacon}, R., {Bureau}, M.,
  {Cappellari}, M., {Davies}, R.~L., {de Zeeuw}, T., {Falc{\'o}n-Barroso}, J.,
  {Krajnovi{\'c}}, D., {Kuntschner}, H., {Peletier}, R.~F., and {Sarzi}, M.:
  2006,
\newblock {\em \mnras} {\bf 373}, 906

\end{thebibliography}
%


\printindex
\end{document}